\documentclass[a4paper,11pt]{article}
\usepackage{a4wide} 
\usepackage{graphics}

\title{Singularity spectra of rough growing surfaces from wavelet analysis}
\author{M. Ahr\footnotemark[1], M. Biehl \\
Institut f\"{u}r Theoretische Physik \\
Julius-Maximilians-Universit\"{a}t 
W\"{u}rzburg\\
 Am Hubland, 97074 W\"{u}rzburg, Germany }
\begin{document}
\maketitle
\setlength{\unitlength}{\textwidth}
\begin{abstract}
We apply the {\it wavelet transform modulus maxima} (WTMM) 
method \cite{adr99} to the analysis of simulated 
MBE - grown surfaces. In contrast to the structure function approach commonly
used in the literature, this new method permits an investigation of the 
complete singularity spectrum. We focus on a kinetic Monte-Carlo model 
with Arrhenius dynamics, 
which in particular takes into consideration the process of 
thermally activated desorption of particles.
  We find a wide spectrum of H\"{o}lder exponents, which reflects the 
multiaffine surface morphology. Although our choice of parameters 
yields small desorption rates ($< 3 \%$), we observe a dramatic change in
the singularity spectrum, which is shifted towards smaller H\"{o}lder 
exponents.    
Our results offer a mathematical foundation of anomalous scaling: 
We identify the global exponent $\alpha_{g}$ with the H\"{o}lder 
exponent which maximizes the singularity spectrum. 
\end{abstract}

\footnotetext[1]{e-mail:{\ttfamily ahr@physik.uni-wuerzburg.de } }
\section{Introduction}
Inspired by the great technological importance of epitaxial crystal growth, 
the past decade has raised much theoretical research in the subject of 
{\em kinetic roughening} of surfaces during growth. The investigation
of this effect, which 
is undesirable in practical applications, promises deep insight 
into statistical physics far from thermal equilibrium, 
see e.g. \cite{bar95} for an overview.
We focus on a full-diffusion Monte-Carlo model of homoepitaxial growth 
of a hypothetical material with simple cubic lattice structure 
under {\em solid on solid} conditions, i.e. the effects of overhangs 
and displacements are being neglected. Then, the crystal can 
be described by a two-dimensional array of integers which denote the 
height $f(\vec{x})$ of the surface.  
On each site, new particles are deposited with a rate $r_{a}$. Particles 
on the surface are hopping to nearest neighbour sites with Arrhenius 
rates $\nu_{0} \exp(-(E_{b} + n E_{n})/(k_{b}T))$, where $E_{b}$ 
and $E_{n}$ 
are the binding energies of a particle to the substrate and to 
its $n$ nearest neighbours. $\nu_{0}$ is the attempt frequency, and 
$k_{b} T$ has its usual meaning. 
In contrast to earlier investigations of similar models \cite{dslkg96},
we permit
the {\em desorption} of particles from the surface with rates
$\nu_{0} \exp(-(E_{d} + n E_{n})/(k_{b}T))$, where $E_{d} > E_{b}$.  

The aim of this publication is twofold: We will first 
discuss the advantages of the wavelet analysis compared to the {\em structure 
function} (SF) approach, which has to date solely 
been used in the investigation of multiaffine surfaces. 
Then, we will apply this formalism to investigate the influence of desorption
on kinetic roughening. We conclude with some remarks on the relevance 
of universality classes for our results.   

\section{Scaling concepts}
The standard approach 
of {\em dynamic scaling} \cite{bar95}
assumes that the statistical properties of a 
growing surface before saturation remain invariant under a simultanous 
transformation of spatial extension $\vec{x}$, heigth $f(\vec{x})$ 
and time $t$,
\begin{equation}
\label{dynamisch}
\vec{x} \rightarrow b \vec{x} = \vec{x}'; \ \
f  \rightarrow b^{\alpha} f = f'; \ \
t  \rightarrow b^{z} t = t';
\end{equation}
where $b$ is an arbitrary positive constant. This implies, that a part of 
the surface smaller than the {\em correlation length} 
$\xi(t) \sim t^{1/z}$ can be regarded as {\em self-affine} with Hurst 
exponent $\alpha$.  
A popular method of measuring $\alpha$ uses heigth-heigth correlation
functions of (theoretically) arbitrary order $q$: 
\begin{equation}
G(q, \vec{l}, t) := \left< \left| f(\vec{x}, t) - f(\vec{x} + \vec{l}, t)
                 \right|^{q} 
                 \right>_{\vec{x}} \sim l^{q \alpha} 
                 g(l/\xi(t)), 
\label{gg}
\end{equation}
where $g(x) \rightarrow \mbox{const.}$ for $x \rightarrow 0$ and 
$g(x) \rightarrow \mbox{const.} \cdot x^{- q \alpha}  $ 
for $x \rightarrow \infty$. 
In practice, $q = 2$ is the most common choice.

In principle, there are two different ways to measure $\alpha$: The {\em 
local} approach determines $\alpha$ from the initial 
slope of $\ln(G(q, \vec{l}, t))$ 
versus $\ln(l)$ for small $l$. The {\em global} approach analyzes 
the dependence of the {\em surface width} 
$w = \sqrt{\left< (f(\vec{x}, t) - \left< f(\vec{x}, t) \right> )^{2} 
\right>_{\vec{x}}}$ 
in the saturation regime on the system size $N$: 
$w_{sat}(N) \sim N^{\alpha_{g}}$.
Before saturation, the surface width increases like $w \sim t^{\beta}$, 
where $\beta = \alpha/z$. 
An alternative which avoids the simulation of different system sizes 
uses the 
complete functional dependence of equation \ref{gg}: $\alpha_{g}$ and $z$ are 
chosen such that the curves of $G(2, \vec{l}, t)/l^{2 \alpha_{g}}$ 
versus $l/t^{1/z}$ collapse on a unique function $g$ within a large range
of $t$ and $l$. 

However, a careful analysis of simulation data
\cite{bbjkvz92,k94,dsp97,dslkg96}
has shown, that several models of epitaxial growth show significant deviations
from this simple picture. First, one obtains different values of $\alpha$ 
from the local than from the global approach, a phenomenon which is
called {\em anomalous scaling}. Second, one often finds {\em multiscaling}: 
height-height correlation functions of different order yield a hirarchy 
of $q$-dependent exponents $\alpha(q)$, when determined from the 
initial power-law behaviour of $G(q, \vec{l}, t)$. 

These observations can be interpreted within the mathematical framework of 
{\em multifractality}: The {\em H\"{o}lder exponent} \cite{adr99,bar95,bma92,mba93} 
$h(\vec{x_{0}})$ of a 
function $f$ at $\vec{x}_{0}$ is defined as the largest exponent such that 
there exists a polynomial of order $n < h(\vec{x}_{0})$ and a constant 
$C$ which yield $|f(\vec{x}) - P_{n}(\vec{x} - \vec{x}_{0})| \leq 
C |\vec{x} - \vec{x}_{0}|^{h(\vec{x}_{0})}$ in the neighbourhood of 
$\vec{x}_{0}$. 
The H\"{o}lder exponent is a {\em local} counterpart of the Hurst 
exponent: a self-affine function with Hurst exponent $\alpha$ has 
$h(\vec{x}) = \alpha$ everywhere. 
However, in the case of a {\em multiaffine} function different 
points $\vec{x}$ might be characterized by different H\"{o}lder exponents.
This general case is characterized by the {\em singularity spectrum} 
$D(h)$, which denotes the Hausdorff dimension of the set of points, where 
$h$ is the H\"{o}lder exponent of $f$.

\section{The wavelet approach to multifractality}
There is a deep analogy between multifractality and thermodynamics
\cite{bar95,s90,v92}, where the scaling exponents play the role of energy, 
the singularity spectrum corresponds to entropy, and $q$ plays the role of 
inverse temperature. 
So, theoretically $D(h)$ might be calculated via a Legendre transform of 
$\alpha(q)$: $D(h) = \mbox{min}_{q} (q h - q \alpha(q) + 2) $
\cite{adr99,bma92,mba93}, a method which has been called 
{\em structure function} (SF) 
approach. However, its practical application raises fundamental 
difficulties: First, to obtain the complete singularity spectrum, one 
needs $\alpha(q)$ for positive {\em and} negative $q$. But
as $|f(\vec{x}, t) - f(\vec{x} + \vec{l}, t)|$ might become 
zero, $G(q, \vec{x}, t)$ is in principle undefined for $q < 0$. Therefore, only
the left, ascending part of $D(h)$ is accessible to this method.   
Additionally, the results of the SF method can  
easily be corrupted by polynomial trends in $f(\vec{x})$ \cite{bma92}. 
It might be due to these difficulties, that - to our knowledge - no attempt 
to determine the singularity spectrum of {\em growing surfaces}
from $\alpha(q)$ has ever been made. 
Although it has been argued that the $\alpha(q)$ collapse onto a single 
$\alpha$ in the limit $t \rightarrow \infty $,
which characterizes the asymptotic universality class of the model 
\cite{dslkg96,dsp97,dkdds97},
we are convinced that deeper insight into fractal growth 
on experimentally relevant {\em finite } 
timescales can be gained from a detailed knowledge of 
the $D(h)$ spectrum. 

To this end, we 
follow the strategy suggested by Arn\'{e}odo et. al \cite{adr99,wave,bma92}, 
which circumvents the problems of the SF approach 
and permits a reliable measurement of the complete $D(h)$. 
Mathematically, the wavelet transform of a function $f(\vec{x})$ of two 
variables is defined as its convolution with the complex conjugate of 
the wavelet $\psi$, which is {\em dilated} with the {\em scale}
 $a$ and rotated by an angle $\theta$ \cite{wave}: 
\begin{equation}
T_{\psi}[f] (\vec{b}, \theta, a) = C_{\psi}^{-1/2} a^{-2} \int d^{2}x
\; \psi^{*}(a^{-1} \mathbf{R_{-\theta}} (\vec{x} - \vec{b}) ) f(\vec{x}). 
\label{wgen}
\end{equation} 
Here $\mathbf{R_{\theta}}$ is the usual 2-dimensional 
rotation matrix, and $C_{\psi} = (2 \pi)^{2} \int d^{2}k |\vec{k}|^{-2} 
|\hat{\psi}(\vec{k})|^{2}$ is a normalization constant, whose 
existence  requires square integrability of the  wavelet $\psi(\vec{x})$ in 
fourier space. Apart from this constraint, the wavelet can (in principle) be
an arbitrary complex-valued function.  
Introducing the wavelet $\psi_{\delta}(\vec{x}) 
= \delta(\vec{x}) - \delta(\vec{x} + \vec{n})$, where $\vec{n}$ is an 
arbitrary unit vector, one obtains easily
\begin{equation}
T_{\psi_{\delta}}[f](\vec{b}, \theta, a) = C_{\psi_{\delta}}^{-1/2} \left[ 
f(\vec{b}) - f(\vec{b} + a \mathbf{R_{\theta}} \vec{n}) \right] ~ \Rightarrow 
~ \int d^{2}b \left| T_{\psi_{\delta}}[f](\vec{b}, \theta, a)\right|^{q} \propto 
G(q, a \mathbf{R_{\theta}} \vec{n}).
\label{tg}
\end{equation} 
Consequently, a calculation of the moments of the wavelet transform of the surface
yields the SF approach as a special case. To avoid its weaknesses, two major 
improvements are necessary:

First, we use a class of wavelets 
with a greater number of {\em vanishing moments} $n_{\vec{\Psi}}$  than 
$\psi_{\delta}(\vec{x})$. This increases the range of accessible H\"older 
exponents and improves the insensitivity to polynomial trends in $f(\vec{x})$. 
We introduce a two-component version of the wavelet transform 
\begin{equation}
\vec{T}_{\vec{\Psi}}[f] (\vec{b}, a) = \frac{1}{a^{2}} \int d^{2}x \left(
\begin{array}{c}
\Psi_{1}(a^{-1} (\vec{x} - \vec{b})) \\
\Psi_{2}(a^{-1} (\vec{x} - \vec{b}))
\end{array}
\right)
f(\vec{x}) \; , 
\label{wspec}
\end{equation}
where the analyzing wavelets $\Psi_{1}$, $\Psi_{2}$ are defined as partial 
derivatives 
of a radially symmetrical convolution function $\Phi(\vec{x})$: 
$\Psi_{1}(\vec{x}) = \partial \Phi / \partial x$, 
$\Psi_{2}(\vec{x}) = \partial \Phi / \partial y$. Then 
$\vec{T}_{\vec{\Psi}}[f](\vec{b}, a)$ can be written as the gradient 
of $f(\vec{x})$, smoothed with a filter $\Phi$ w.r.t. $\vec{b}$.   
This definition
becomes a special case of equation \ref{wgen}, when multiplied with
$\vec{n}_{\theta} = (\cos(\theta), \sin(\theta))^{\top}$, yet allows for an 
easier numerical computation\footnote{For simplicity, the irrelevant 
constant $C_{\Psi}$ has been omitted}.
For example, $\Phi$ can be a gaussian, where $n_{\vec{\Psi}} = 1$, or 
$\Phi_{1}(\vec{x}) = (2 - \vec{x}^{2}) \exp(-\vec{x}^{2}/2)$, which has two 
vanishing moments.   

Second, the integration over $\vec{b}$ in equation 
\ref{tg} is undefined for $q < 0$, since the wavelet coefficients 
might become zero. The basic idea is to replace it with a discrete summation 
over an appropriate partition of the wavelet transform which obtains
nonzero values only, 
but preserves the relevant information on the H\"older 
regularity of $f(\vec{x})$. In the following, we will give a brief outline of 
the rather involved algorithm and refer the reader to \cite{adr99,wave,bma92}
for more details and a mathematical proof.    
The {\em wavelet transform modulus maxima} (WTMM) are defined as 
local maxima of the modulus 
$M_{\vec{\Psi}}[f](\vec{b}, a) := |\vec{T}_{\vec{\Psi}}[f](\vec{b}, a)|$
in the direction of 
$\vec{T}_{\vec{\Psi}}[f](\vec{b}, a)$ for fixed $a$. 
These WTMM lie on connected curves, which trace
structures of size $\sim a$ on the surface. The 
strength of each is characterized by the {\em maximal} value 
of $M_{\vec{\Psi}}[f](\vec{b}, a)$ along the line, the so-called 
{\em wavelet transform modulus maxima maximum} (WTMMM) \cite{adr99}.  
While proceeding from large to small $a$, successively smaller structures 
are resolved. Connecting the WTMMM at different scales yields the set 
$\mathcal{L}$ of maxima lines $l$, which lead to the locations of the 
singularities of $f(\vec{x})$ in the limit $a \rightarrow 0$.  
The partition functions 
\begin{equation}
Z(q, a) = \sum_{l \in {\mathcal L}(a)} \left(
\sup_{(\vec{b}, a') \in l, 
a' \leq  a} M_{\vec{\Psi}}[f](\vec{b}, a')
\right)^{q} \sim a^{\tau(q)} \; \mbox{for} \; a \rightarrow 0
\end{equation}
are defined on the subset $\mathcal{L}(a)$ of lines which cross the scale $a$.
 From the analogy between the multifractal formalism and 
thermodynamics, $D(h)$ is calculated via a legendre transform of the 
exponents $\tau(q)$, which characterize the scaling behaviour of $Z(q, a)$  
on small scales $a$: $D(h) = \mbox{min}_{q} (q h - \tau(q))$.   
Additionally, $\tau(q)$ itself has a physical meaning for some $q$: 
$-\tau(0)$ is the fractal dimension of the set of points where 
$h(\vec{x}) < \infty$, while the fractal dimension of the surface 
$f(\vec{x})$ itself equals $\mbox{max} (2, 1 - \tau(1))$. 

\section{Results}
\begin{figure}
\begin{center}
\begin{picture}(1,0.4)(0,0)
\put(0,0){\resizebox{0.48\textwidth}{!}{\includegraphics{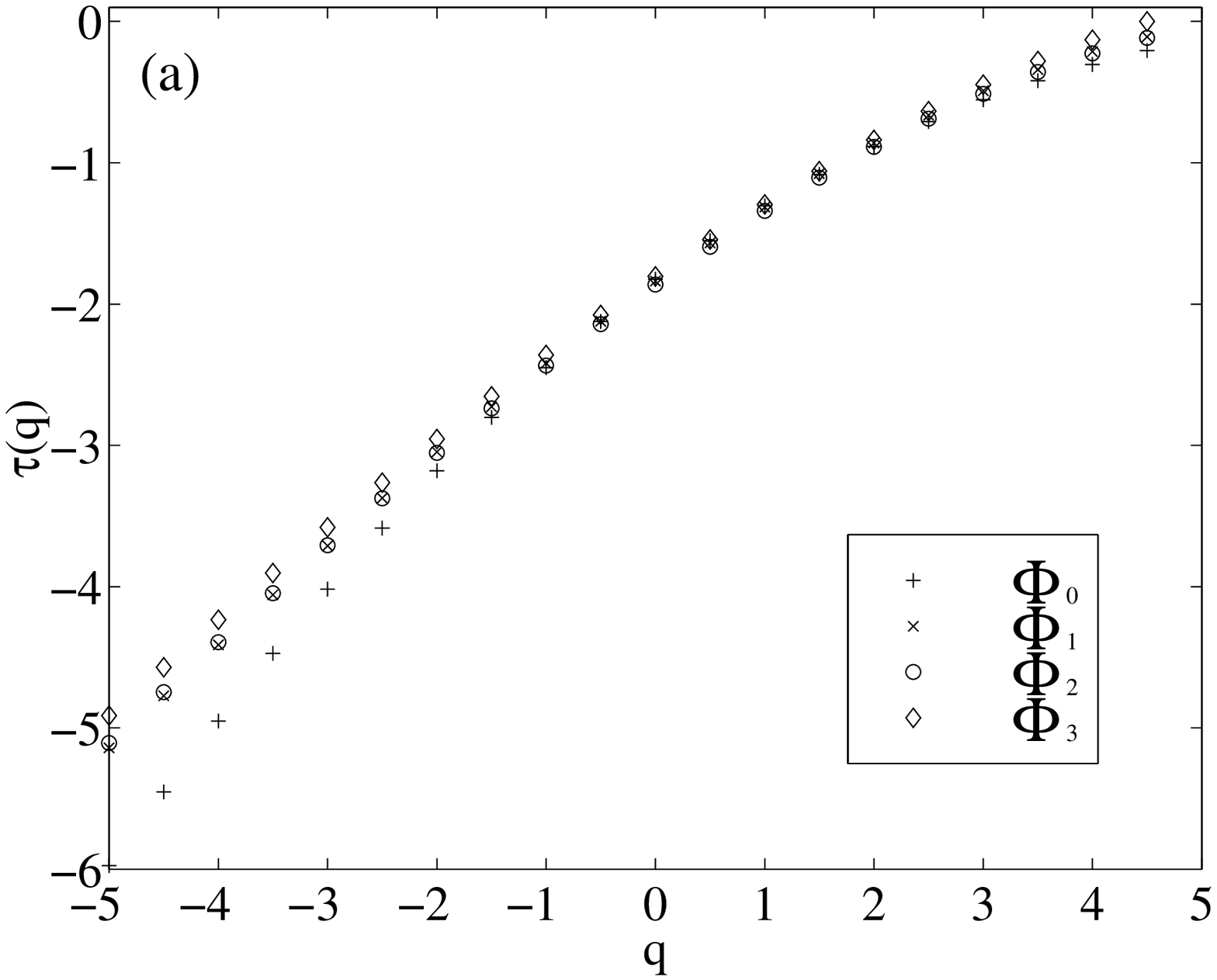}}}
\put(0.52,0){\resizebox{0.48\textwidth}{!}{\includegraphics{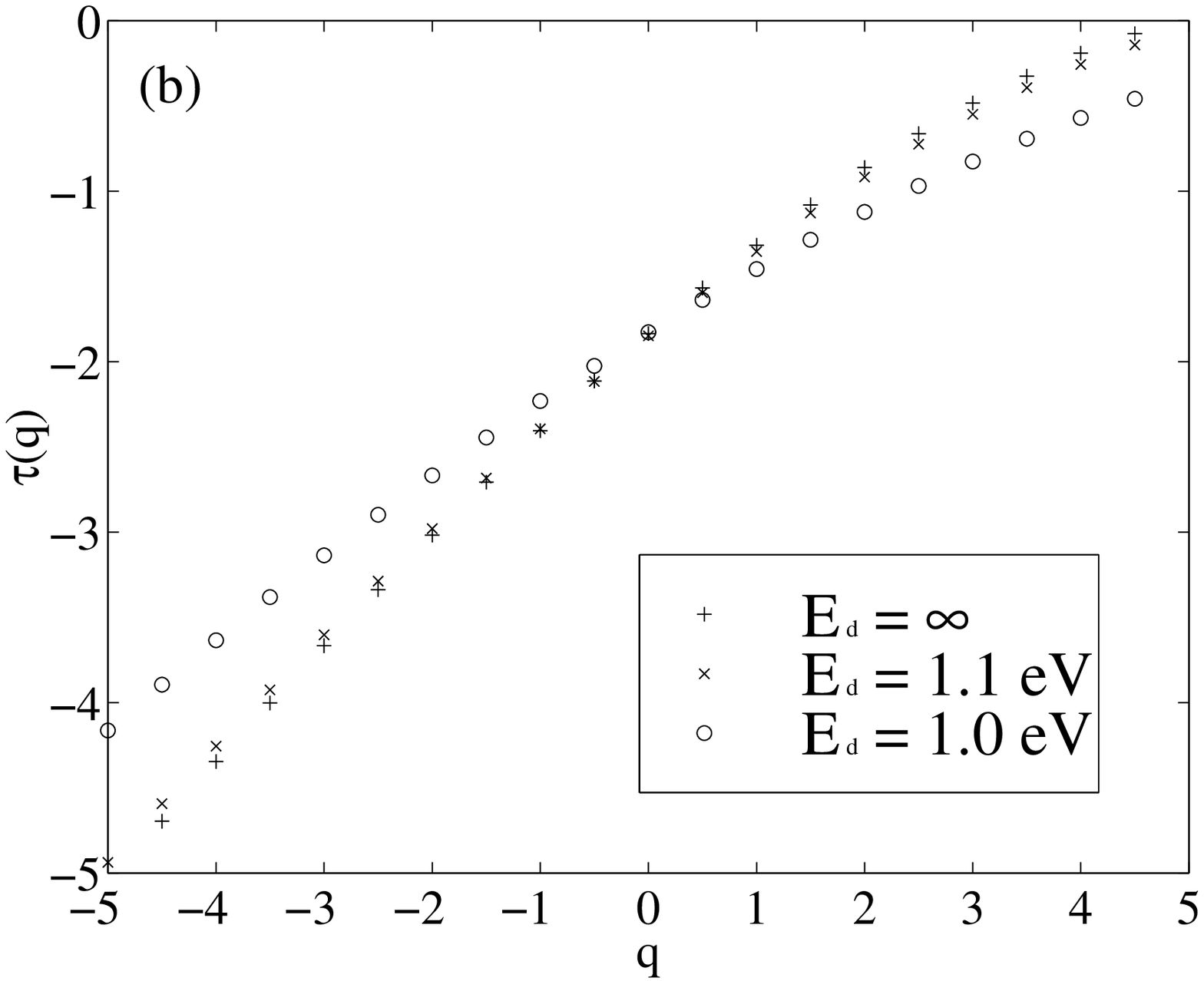}}}
\end{picture}
\end{center}
\caption{Panel a: $\tau(q)$ in the $E_{d} = \infty$ - model as 
obtained from investigations with different wavelets. Note the deviations of 
the data obtained with the gaussian function $\Phi_{0}$.
Panel b: $\tau(q)$ for models with different activation energies $E_{d}$ 
of desorption. All data have been obtained from surfaces after 
$2 \cdot 10^{4} s$ 
of simulated time. Sizes of errorbars are on the order of symbol sizes.}
\label{eins}
\end{figure}
In our simulations, we choose the parameters $\nu_{0} = 10^{12} / s$, 
$E_{b} = 0.9 \; \mbox{eV}$ and  $E_{n} = 0.25 \; \mbox{eV}$,
and a temperature $T = 450 \mbox{K}$. 
To study the influence of desorption, we consider three models with 
different activation energies $E_{d}$: in model A desorption is 
forbidden, i.e. $E_{d} = \infty$. Models B and C have 
$E_{d} = 1.1 \; \mbox{eV}$ and $E_{d} = 1.0 \; \mbox{eV}$. 
We simulate the deposition of $2 \cdot 10^{4}$ monolayers at a growth rate 
of one monolayer per second on a lattice of $N \times N$ unit cells 
using periodic boundary conditions, our standard value being 
$N = 512$. To check for finite size effects, we have also simulated 
$N = 256$.   
In all presented results averages over 6 independent simulation runs have 
been performed.
Although we have used an optimized algorithm, these simulations consumed 
several weeks of CPU time on our workstation cluster. 

First, we have checked our results for artifacts resulting 
from properties of the  analyzing wavelet rather than from 
the analyzed surface by using different 
convolution functions $\Phi_{n}$: $\Phi_{0}$ is the gaussian function, 
$\Phi_{n}, n \geq 1$ are products of gaussians and 
polynomials, which have been chosen in a way that the first $n$ 
moments vanish. Then, the analyzing wavelets have 
$n_{\vec{\Psi}_{n}} = n + 1$ vanishing moments. 
We find (figure \ref{eins} a), 
that the $\tau(q)$-curve obtained with $\Phi_{0}$ deviates 
significantly from those obtained with $\Phi_{1}$, $\Phi_{2}$ and $\Phi_{3}$. 
The latter agree apart from small differences which are mainly 
due to the discrete sampling of the wavelet in the numerical implementation
of the algorithm.
This is explained by the theoretical result \cite{mba93} that  
$d \tau(q) / dq = n_{\vec{\Psi}}$ for $q < q_{crit.} < 0$ 
if the number of vanishing moments of the analyzing wavelet is too small.
Consequently, the agreement of the other curves proves their 
physical relevance.

Figure \ref{eins} b shows averages of $\tau(q)$ curves obtained with 
the convolution functions $\Phi_{1}$, $\Phi_{2}$, $\Phi_{3}$ 
from surfaces after  $2 \cdot 10^{4} s$ 
of growth on an initially flat substrate.
For all our models, their nonlinear behaviour reflects the multiaffine surface 
morphology. From the fact that these curves are reproduced 
within statistical errors in simulations with $N = 256$, we conclude 
that finite size effects can be neglected. 
Clearly, desorption reduces the slope of $\tau(q)$, although only a small 
fraction of the incoming particles is desorbed: $0.18 \%$ in model B
and $2.57 \%$ in model C with slightly higher values at earlier times. 
\begin{figure}
\begin{center}
\begin{picture}(1,0.4)(0,0)
\put(0,0){\resizebox{0.48\textwidth}{!}{\includegraphics{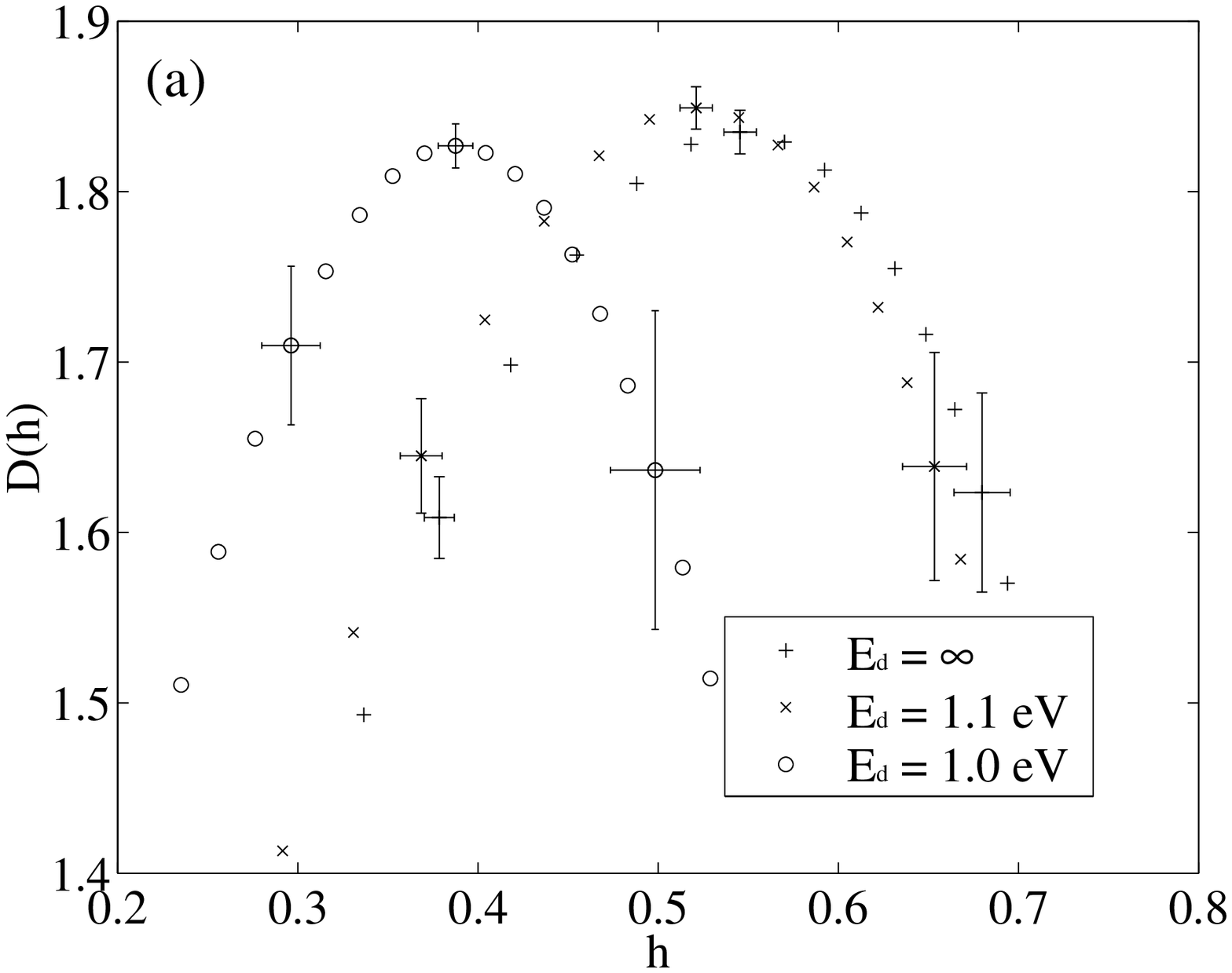}}}
\put(0.52,0){\resizebox{0.48\textwidth}{!}{\includegraphics{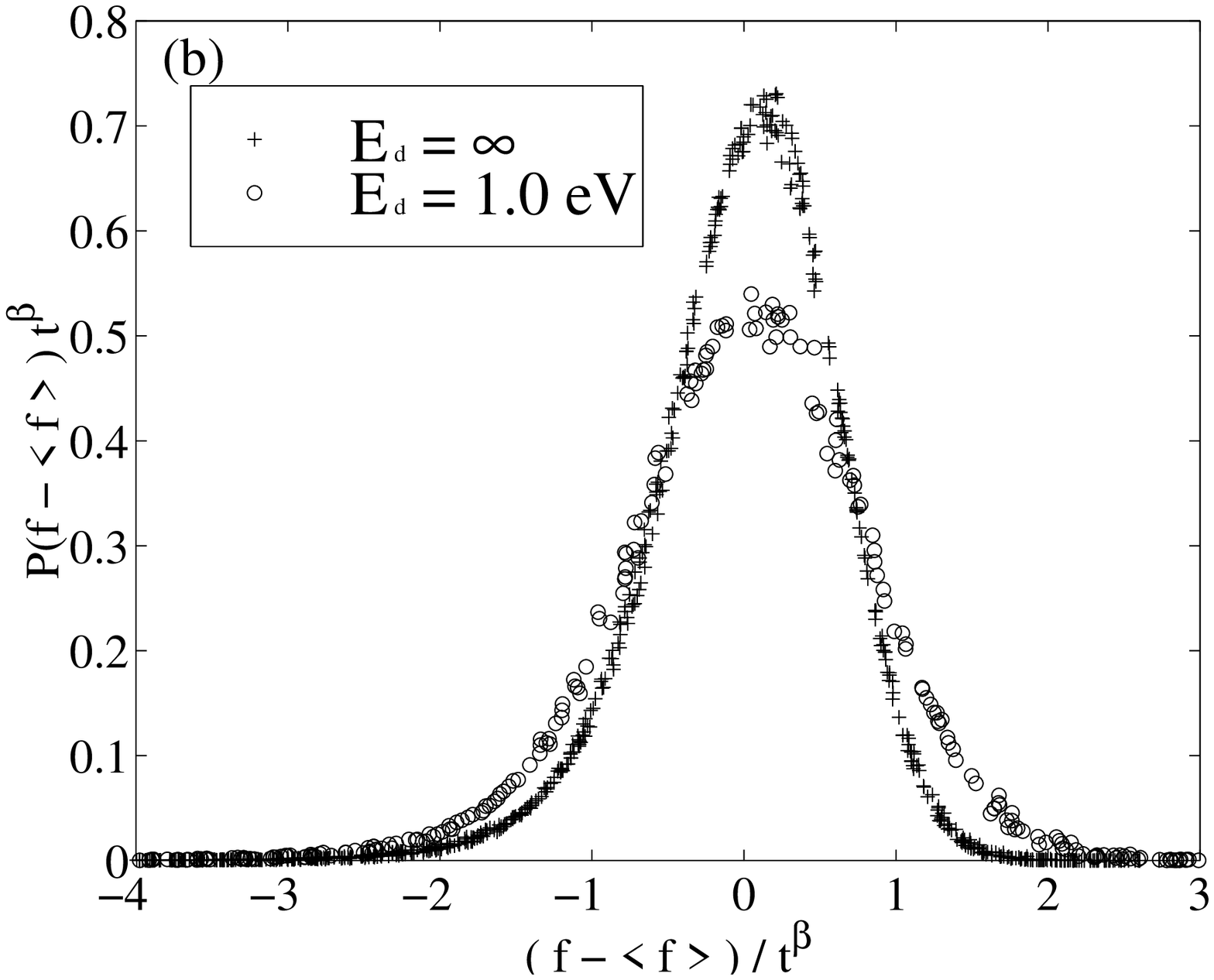}}}
\end{picture}
\end{center}
\caption{Panel a: singularity spectra obtained from a Legendre transform 
of the data in fig. \ref{eins} b. Panel b: Data collaps 
of rescaled PDFs of surface heights for times between 
$150 s$ and $2\cdot 10^{4} s$ (model A) respectively
$150 s$ and $7500 s$ (model C). 
We have used $\beta = 0.188$ for model A with $E_{d} = \infty$ and 
$\beta = 0.109$ for model C ($E_{d} = 1.0 eV$). Time is measured in 
seconds.}
\label{zwei}
\end{figure}
The corresponding singularity spectra are shown in figure \ref{zwei} a.
They have a typical shape whose descending part seems to be symmetrical 
to the ascending part and which changes at most slightly, while
the whole spectra are shifted towards smaller H\"{o}lder exponents as 
desorption becomes more important. 
We emphasize that we find no evidence for a {\em time dependence} 
of the singularity spectra within the range 
$9700 s \leq t \leq 2 \cdot 10^{4} s$, so that our results do {\em not}
support the idea of an asymptotic regime characterized by a single 
exponent $\alpha$. However, the accessible time range of computer 
simulations is limited, so that we cannot finally disprove the existence of
such a regime. 

The multifractal formalism has replaced the unique scaling exponent 
$\alpha$ of {\em spatial extension} 
in the simple picture of dynamic scaling (equation \ref{dynamisch})
with a wide spectrum of H\"{o}lder exponents. By analogy, one might 
find it necessary to replace the scaling exponent $\beta$ 
with a distribution of {\em temporal}
counterparts of $h$. 
To answer this question, we investigate the propability 
distribution function (PDF) $P(f - \left<f\right>, t)$ of surface heights.
Dynamical scale invariance with a single $\beta$ demands that  
\begin{equation}
P(f - \left<f\right>, t) = 
\tilde{P}\left(\frac{f - \left<f\right>}{t^{\beta}} \right)
\frac{1}{t^{\beta}} \; ,
\label{bkollaps}
\end{equation}
i.e. the rescaled PDFs $P t^{\beta}$ should collapse onto 
a single function $\tilde{P}$ when plotted as a function of 
$(f - \left<f\right>)/t^{\beta}$ within a large time range.  

We measure $\beta$ from the increase of the surface width 
with time, which follows a power
law for $t \geq 150 s$ in models A and B ($\beta_{A} = 0.19 \pm 0.01$, 
$\beta_{B} = 0.17 \pm 0.01$) respectively $ 150 s < t < 7500 s$ in model 
C ($\beta_{C} = 0.11 \pm 0.01$), which then starts to approach the 
final {\em saturation regime}.
 The high quality of the data collapse of the PDFs shown in figure
\ref{zwei} b proves that the scaling form \ref{bkollaps} holds, showing 
that a {\em single} exponent describes the scaling behaviour of 
$P(f - \left<f\right>, t)$. This parallels the finding of Krug in \cite{k94} 
for the one-dimensional Das Sarma-Tamborenea model.
 
Finally, the WTMM method, which is a precise tool to investigate {\em local} 
scaling properties of surfaces might help to get some insight into the 
phenomenon of {\em anomalous scaling}. 
The conventional picture \cite{lr96,lrc97,l99} notes the 
difference between the global $\alpha_{g}$ and a ``local $\alpha$'' which 
is determined from the power-law behaviour of $G(2, \vec{l}, t)$ for 
small $l$, and, within the multifractal formalism, simply 
corresponds to a H\"{o}lder exponent
on the ascending part of the singularity spectrum. 
We have determined the 
{\em global} scaling exponents $\alpha_{g}$ and $z$ from 
the data collapse of the scaled height-height correlation function 
$G(2, \vec{l}, t)$ and find agreement within statistical errors 
between $\alpha_{g}$ and that value of the H\"{o}lder exponent 
$h_{m}$ which {\em maxmizes} $D(h)$ (table \ref{tabelle}). 
This empirical result can be explained with a saddle-point argument:
We calculate the surface width 
\begin{equation}
w^{2} = \frac{1}{N^{2}} \int d^{2} x (f(\vec{x}) - \left<f\right>)^{2} 
= \frac{1}{N^{2}} \int d\tilde{h} \underbrace{
\int d^{2}x \; \delta(\tilde{h} - h(\vec{x}))
(f(\vec{x}) - \left<f\right>)^{2}}_{I(\tilde{h})} \; . 
\end{equation}
Since $I(\tilde{h})$ grows like $N^{D(\tilde{h})}$ with the system size, 
in large systems the integral over $\tilde{h}$ will be dominated by 
$I(h_{m})$. 
That means, that $w$ and therefore the global scaling properties of the 
surface are governed by the subset of points, which has the greatest 
fractal dimension. Consequently, the surface will behave like a self-affine 
surface with Hurst exponent $h_{m}$ on lenghtscales comparable to the 
system size.  

\section{Conclusions}
\begin{table}
\begin{center}
\begin{tabular}[t]{|l|c|c|c|c|c|c|}
\hline
Model & $p_{d}$ & $\beta$       &$h_{m}$        &$\alpha_{g}$& $z_{g}$ & $D_{f}$ \\ \hline 
A     & 0       &0.19 $\pm$ 0.01&0.54 $\pm$ 0.01& 0.55       & 2.9     & 2.32 $\pm$ 0.01 \\
B     & 0.18 \% &0.17 $\pm$ 0.01&0.52 $\pm$ 0.01& 0.51       & 3.3     & 2.35 $\pm$ 0.01 \\
C     & 2.57 \% &0.11 $\pm$ 0.01&0.38 $\pm$ 0.01& 0.39       & 3.5     & 2.45 $\pm$ 0.02 \\
\hline
\end{tabular}
\end{center}
\caption{Simulation results: $p_{d}$ is the fraction of particles which desorbes, 
$\beta$ is the scaling exponent of the PDF of surface heights, $h_{m}$ the 
H\"{o}lder exponent which maximizes $D(h)$, $\alpha_{g}$ and $z_{g}$ are the 
global scaling exponents of $G(2, \vec{l}, t)$, $D_{f}$ is the fractal 
dimension of the surface. }
\label{tabelle}
\end{table}
Table \ref{tabelle} summarizes our results. Model A without 
desorption reviews the results in \cite{dslkg96}, which have been 
obtained with slightly different activation energies on smaller systems 
and shorter timescales. Models B and C show, that desorption is
an important process, which, although it affects only a small fraction
of the adsorbed particles, must not be neglected, since it alters the
scaling properties of the surfaces by reducing $\beta$ and by shifting 
the singularity spectrum towards smaller H\"{o}lder exponents. Since 
the scaling behaviour depends strongly on the height of the 
energy barrier of desorption, and the singularity spectra have no measurable
tendency to narrow with time, our results can not be used to make 
any decision on the aymptotic universality class of the investigated model.
However, they show, that the paradigm of a few universality classes 
characterized by a {\em small} number of exponents, which are 
{\em independent} on details of the model, is {\em not} adequate to catch the 
features of kinetic roughening on experimentally relevant timescales of a 
few hours of growth. 

We are convinced, that the application of new mathematical tools 
like the wavelet analysis will help to find a better description of 
fractal growth phenomena in the future.

We thank A. Arn\'{e}odo and J. M. L\'{o}pez for providing us recent 
preprints before publication and A. Freking for a critical reading 
of the manuscript.

\end{document}